# Publish or Perish: A population-dynamics model of the corruption of peer review

Shyam Ranganathan,

Department of Philosophy,

York Centre for Asian Research,

York University

## Abstract

Peer review is commonly treated as a mechanism of quality control. This article examines what happens when participants are split between two groups. *Explicators* assess whether submissions present logic-based arguments (whether deductive, inductive or abductive), independently of whether they agree with the reasons or the conclusion. *Interpreters* assess submissions according to their outlook. Their employment of logic is opportunistic, as a means of presenting their own views, but not as a means of assessing others' views. Using a deterministic population-dynamics model of iterative games of "Publish or Perish," this article shows how this asymmetry permits interpreters to increase their population share even when they begin as a small minority. It compares fixed and survivor-drawn reviewer panels, alternative replenishment rules, hostile explicator responses, generational differences in ratios, and different panel sizes. Under survivor-drawn review, explicator hostility to interpreters produces an unstable equilibrium at an explicator share of $q* = 2 - \sqrt{3} \approx 0.2679$: below this threshold, interpreters can survive without explicator cooperation and continue toward dominance. The threshold is not universal, however. Exogenous replenishment in favour of explicators can eliminate the point of no return, and sufficiently lagged, generational replenishment can permit explicators to recover after it has apparently been crossed. The model therefore identifies not only a mechanism by which peer review becomes corrupted, but also the institutional conditions determining whether that corruption becomes self-sustaining or reversible.

# Introduction

Peer review is ordinarily justified as a procedure for assessing the quality of scholarly contributions. That justification presupposes that reviewers can distinguish the acceptability of an argument from their agreement with its conclusion. A reviewer may believe that a paper is mistaken while nevertheless recognizing that it advances a valid argument, identifies a genuine alternative, or clarifies the structure of an existing controversy. Without this distinction, peer review ceases to test whether a submission contributes to inquiry and instead tests whether it conforms to the outlooks already represented among reviewers.

This article models the consequences of replacing the first standard with the second. It distinguishes between two approaches to assessment: *explication* and *interpretation*. Explication renders reasons and conclusions explicit through deductive, inductive, or abductive reasoning. Explicators assess whether an argument supports its advertised conclusion and contributes to an understanding of the available positions in a controversy. Because validity and cogency do not depend upon agreement with a conclusion, explicators can approve submissions whose conclusions they reject.

Interpretation operates differently. Interpreters assess a submission by placing it within an outlook, belief, or propositional attitude and asking whether it accords with that outlook. An interpreter may employ the outward form of argument and may therefore produce work that satisfies minimal explicatory scrutiny. But argumentative form does not determine how the interpreter assesses the work of others. When acting as a reviewer, the interpreter rejects submissions that conflict with their outlook, regardless of whether those submissions make an explicable contribution to a controversy.

This produces a structural asymmetry. Explicators can approve interpreter submissions when those submissions are formulated as arguments. Interpreters, by contrast, approve only those explicator submissions with which they agree. The resulting advantage does not depend upon interpreters producing better arguments, being initially more numerous, or coordinating their behaviour. It follows from the interaction between two different standards of acceptance. One population extends a procedure-independent form of consideration that the other population does not reciprocate. One population assesses everything substantively. The other does not.

The “Publish or Perish” game introduced in earlier work (Ranganathan 2024, 2026) represents this asymmetry as a repeated selection process. Authors survive a round only if their submissions receive unanimous reviewer approval, and only survivors remain available as authors or, in some versions of the game, as reviewers in subsequent rounds. Under the model’s initial assumptions, explicators approve submissions that pass minimal argumentative scrutiny. Interpreters can meet that standard in presenting their own work, do not eliminate one another, and reject explicator submissions whenever they disagree with them. For simplicity, half of explicator submissions are assumed to agree with interpreters' outlook.

The initial question is whether this selective veto is sufficient to change the composition of the population. It is. With one fixed explicator and one fixed interpreter reviewing each submission, interpreters survive every round while only half of explicators do. When reviewers are instead drawn from the previous round's survivors, the decline of explicators is initially slower because some explicator submissions receive two explicator reviewers. Nevertheless, the same asymmetry alters the composition of later reviewer panels, which in turn intensifies the asymmetry. Even when interpreters begin as only 10% of the population, repeated selection eventually makes them the majority.

The more important question is whether interpreter dominance ever becomes independent of continuing explicator cooperation. Interpreters' employment of logic is opportunistic: they can present their positions in argumentative form, which allows their submissions to pass explicator scrutiny, but they do not reciprocate that standard when reviewing others. But explicators might recognize the asymmetry and respond by rejecting interpreter submissions outright. Is there a distribution beyond which the response is too late?

For two-reviewer panels drawn from the surviving population, the answer is yes. Under explicator hostility to interpreters, an interpreter submission survives only when both reviewers are interpreters, whereas an explicator submission survives before two explicators and, by assumption, half the time before any panel containing an interpreter. Equating the resulting survival rates yields the critical explicator share

$$q^* = 2 - \sqrt{3} \approx 0.2679.$$

This is an unstable equilibrium. Above it, explicators have the higher survival rate and can recover if they cease approving interpreter submissions. Below it — corresponding to an interpreter share above approximately 73.21% — interpreters have the higher survival rate even under universal explicator hostility. At that point, interpreters no longer need explicators to regard their submissions as reasonable. Their numerical control over reviewer selection is sufficient to reproduce their population advantage. This is the model's point of no return.

That threshold is not, however, an invariant feature of every institutional arrangement. The subsequent analysis shows that replenishment must be treated as part of the selection mechanism rather than as a neutral background condition. A fixed 90:10 replenishment of explicators and interpreters continually regenerates the explicator population and can eliminate the point of no return altogether. Replenishment that tracks the current population, by contrast, preserves the same unstable threshold found in the unreplenished survivor model. Replenishment can therefore slow, accelerate, prevent, or consolidate interpreter dominance depending upon how vacancies are filled.

The timing of replenishment matters as well. When replenishment reflects the current population, crossing the threshold produces a stable movement toward interpreter dominance. A short lag in the population data used for replenishment produces temporary reversals without changing the eventual outcome. A sufficiently long lag, however, continues to introduce explicators at a rate reflecting an earlier and more explicator-heavy population. In the two-reviewer version of the game, a lag of three generations is enough to allow explicators to cross back over the equilibrium boundary and eliminate interpreters. An apparent point of no return can therefore be reversible when institutional reproduction responds slowly to changes in the population it reproduces.

Finally, the analysis generalizes the model to panels requiring the unanimous approval of n reviewers. The equilibrium is then determined by

$$2(1-q)^n = 1 + q^n.$$

As panel size increases, the critical explicator share falls and the corresponding interpreter share rises. Larger unanimous panels thus require interpreters to secure a greater majority before they can survive without explicator cooperation. At the same time, simulations suggest that larger panels may become vulnerable to reversal under shorter replenishment lags, although the reported relationship is not monotonic and requires further stability analysis.

The model is deliberately stylized. It does not claim to reproduce every feature of actual peer review or to establish that all disagreement among reviewers is interpretive. Its purpose is to isolate a specific institutional mechanism: what follows when one population assesses contributions by an impersonal argumentative standard while another uses conformity with its own outlook as a selective veto. The results show that corruption need not begin with a majority, explicit coordination, or a formal change in the rules. It can emerge endogenously from asymmetric standards of assessment and then become embedded in the demographic composition of the reviewing population. Whether that outcome becomes irreversible depends not only upon the preferences of individual reviewers, but upon panel formation, population feedback, replenishment, institutional memory, and the timing of resistance.

# Explication vs. interpretation

*Explication* is the practice of rendering explicit reasons and conclusions by means of logic, whether deductive, inductive or abductive. To explicate some option is to use logical validity, the criterion of good deduction, to assess the option's contribution to possible controversies. An argument is logically valid when, if the premises are true, the conclusion has to be true. Explicated options can be assessed independently of whether the assessor happens to share the belief the conclusion leads to. Two people can disagree about whether an explicated conclusion is true while still agreeing on whether the argument for it is valid.

*Interpretation*, by contrast, explains a claim by placing it within a particular outlook, belief, or propositional attitude; it treats agreement with that outlook, rather than logical validity, as the test of acceptability.

While rarely pointed out, these two approaches are mutually incompatible. Valid arguments can be comprised of false propositions we do not believe, and arguments with all true premises and a true conclusion that we believe can fail to be valid.

But moreover, if we adopt interpretation, it would be a complete accident if we ever identified an argument as reasonable, for the *interpreter*'s criterion of acceptability has nothing to do with logic. And we can generalize this as an explanation of informal fallacies as well.

The fallacy of begging the question, assuming the conclusion as a premise, though always logically valid, is an informal fallacy, for in the absence of someone believing the premise, it has nothing going for it. Similarly, the informal fallacy of *ad hominem* arguments, the idea that we should reject what someone has to say on irrelevant grounds of character or personhood, is also something that arises if one is obsessed with certain beliefs of relevance, and not for any formal logical reason.

Interpretation is hence not merely mutually exclusive of explication: it is the rejection of employing logic as the criterion of understanding. And hence, we could treat these as logical contradictories: explication is the rejection of interpretation, and interpretation is the rejection of explication. One cannot do both at the same time.

Because interpretation collapses assessment into agreement with a prior belief, there is no procedure-independent way to adjudicate a disagreement between two interpretations — one either shares the belief or does not.

This distinction is what separates the two kinds of reviewer in the model below.

Explication can be exacting. An *explicator* can, for instance, in assessing a submission ask:

a) Does the submission have a logically valid argument, or perhaps a strong and possibly cogent inductive case for an observation, an inference to the best explanation that explains the topic of controversy?

b) Does the argument support the advertised thesis?

c) Does the paper accurately represent the explicated views of the positions it mentions?

d) Are the grounds for rejecting some alternative advanced on explicatory grounds, or do they reduce to the author's outlook?

e) Does the submission contribute to our appreciation of the possibilities of disagreement and controversy? (This follows in part from logic not being reducible to any particular belief or outlook.)

For the purpose of this analysis, I will assume that explicators employ a minimal standard of scrutiny. This keeps the review at (a) and (b), and allows for some leeway with (c): an inaccuracy in describing some third-party view can be reframed as the articulation of some third-party view that is a foil to the argument. The most exacting standard is (d) as this tracks the difference between an interpreter and an explicator. If an explicator were to enforce (d) and only approve of submissions that reject options on explicatory grounds, then interpreters would be rejected as a matter of course. Interpreters will tend to fail on d) because in their consideration of relevant factors, they will make propositional attitudes (not the propositions they believe) the adjudicating factor.  In practice what this will look like is the “reason” that interpreters give for their assessment is a description of their belief.

The last constraint is supererogatory to some extent. Submissions that succeed according to e) are explicit about the explicatory project. Submissions that fail on d) couldn't possibly succeed on e).

Interpreters, in contrast, only assent to what they agree to. They can fake reasoning by using logic to frame their positions, but they are generally unwilling to tolerate or allow the publication of anything that does not fit with their point of view. In other words an interpreter could use *Modus Ponens*, for instance, to frame their argument but they have no commitment to logical validity as a means of understanding options. And to that extent they undermine the rationale for presenting their argument as a version of Modus Ponens.

A recent famous example of interpretive obstructions in peer review is to be found in the historical reception of Katalin Karikó's work; see (Franzoni, Stephan, and Veugelers 2022; Karikó and Pollock 2023; Nair 2021; Smith and Brilliant 2024). Karikó pioneered mRNA research, and her work was rejected not because it would contribute to an understanding of research controversies, but because, in the assessment of the reviewers, it was not important. Sometimes these assessments were couched in quantitative language, as in the idea that her work only made “incremental” advances. This is an interpretive assessment as it relies upon the assessor's judgment of what counts as incremental. On explicatory grounds even small advances can provide and supplement our understanding of controversies. Indeed on explicatory grounds what might otherwise seem like a small difference might be significant, as it sheds light on the possibilities of disagreements in ways that were not previously understood.

# The game of Publish or Perish and its game-theoretic implications

I introduced the game of Publish or Perish first in a paper (Ranganathan 2024), and then later in my book *Moral Philosophy and De-Colonialism: The Irrationality of Oppression* (Ranganathan 2026).

In its most basic form it assumes that in order to win one round, an author has to receive the approval of two reviewers. The simplest version, version 1 below, assumes that the reviewers are evenly split, through iterations, while only surviving authors can play again.

There I noted that given certain starting assumptions, listed below, interpreters take over publish or perish. Here I consider, further, outcomes if parameters are tweaked. These include: (a) changing the composition of reviewers to reflect survivors of previous rounds, (b) assessing what happens if interpreters are a minority.

But what was left was the question of whether there is a *point of no return*, past which interpreters no longer need explicators to approve of their work to, as a group, survive and dominate.

Answering this question shows us when interpreters no longer have to pretend to be reasonable. The pretense of reasoning was necessary for explicators to approve of their work. To engage in this pretense is to formulate one's own views by way of explicable arguments, but to withhold assessing positions by explicatory means. In other words, this pretense is cosmetic; it has to do with how interpreters present themselves, not in terms of how they assess the data. And to figure out when they no longer need explicators to approve of their work is to figure out the point at which they no longer have to pretend to be concerned about reason. They can hence show themselves to be irrational, and they can also revel in it. They may use the rhetoric of providing reasons, but the 'reasons' are self-reflexive indexicals that pick out their own mental states.

There is certainly much more to be said about the game theoretic aspects of this game and its various versions. But in this article I'm mainly interested in showing that with some basic constraints we can get to a point where peer review is completely corrupted. That is the point at which interpreters don't even have to pretend to be reasonable.

# Beginning assumptions and notation

Let $E_t$ be the number of explicators after round t, $I_t$ the number of interpreters, and $q_t$ the proportion of explicators:

$$q_t = \frac{E_t}{E_t + I_t}$$

The interpreter share is correspondingly $1-q_t = I_t/(E_t+I_t)$; every table below reports this as a percentage in its rightmost column unless noted otherwise.

Each submission requires the unanimous assent of two reviewers.

1. Explicator reviewers apply minimal argumentative scrutiny.
2. Interpreters can formulate their positions in argumentative form that passes explicator scrutiny.
3. Interpreters do not eliminate one another.
4. Interpreter reviewers reject propositions with which they disagree.
5. Half of explicator submissions express propositions with which interpreters agree.
6. Compatibility is a property of the manuscript: if an explicator proposition agrees with interpreters' outlook, all interpreter reviewers accept it.
7. We begin with a fixed number of participants, who are not replenished.

8. After having examined games of attrition, we consider games where the population is replenished.
9. After having examined games with assumption 2, we consider games that begin with 2∗: explicators reject submissions of interpreters outright.
10. Any population is reported to be eliminated if it has less than 1% of the share, or less than 1 participant. In a game involving 100 participants, that means that we recognize that there cannot be a real life fractional participant. *This is merely a reporting convention rather than a rule governing the recurrence.*

**Reviewer-sampling convention.** Scenario 1 is excluded from this convention: its reviewer panel is fixed at one explicator and one interpreter. From Scenario 2 onward, wherever reviewers are drawn from the population, reviewer types are sampled independently according to the relevant population proportions. Accordingly, the panel probabilities are calculated as $q_t^2$, $2q_t(1-q_t)$, and $(1-q_t)^2$, rather than by finite-population sampling without replacement. This convention carries forward only where a later scenario retains population-based reviewer selection.

**On the doxastic convergence of interpreters .**The assumptions of the game, namely 3, require background agreement on the beliefs used by interpreters in adjudication. This does not require any explicit formal convention, agreement, or coordination. Rather, given that interpreters treat their beliefs and other propositional attitudes as dispositive, common enculturation that consists in exposure to the same literature and conventions will produce a uniformity in outlook across interpreters in any enculturation pool. Education systems, and especially the academy, facilitate these enculturation processes. The doxastic convergence of interpreters is hence explained not by coordination, but by the same individual choice to interpret relative to the same enculturation processes. The convergence will appear uniform when interpreters are adjudicating literature in their shared enculturalization space.

# 1. Fixed reviewers; authors are culled

Every submission is reviewed by one explicator and one interpreter. The explicator accepts both types because both satisfy minimal argumentative scrutiny. The interpreter does not eliminate interpreter submissions but accepts only half of explicator submissions.

$$P(\text{explicator survives}) = \frac{1}{2}$$

$$P(\text{interpreter survives}) = 1$$

The recurrence equations are:

$$E_{t+1} = \frac{1}{2}E_t$$

$$I_{t+1} = I_t$$

Starting with E_0 = 100 and I_0 = 100, the general solution is:

$$E_t = 100\left(\frac{1}{2}\right)^t$$

$$I_t = 100$$

| Round | Explicators | Interpreters | Interpreter share |
|---|---|---|---|
| 0 | 100.000 | 100 | 50.00% |
| 1 | 50.000 | 100 | 66.67% |
| 2 | 25.000 | 100 | 80.00% |
| 3 | 12.500 | 100 | 88.89% |
| 4 | 6.250 | 100 | 94.12% |
| 5 | 3.125 | 100 | 96.97% |
| 6 | 1.563 | 100 | 98.46% |
| 7 | 0.781 | 100 | 99.22% |

The explicator population falls below one expected member in round 7. If whole individuals are rounded down after each round:

$$100 \to 50 \to 25 \to 12 \to 6 \to 3 \to 1 \to 0$$

Under that convention, explicators are eliminated in round 7.

# 2. Reviewers are drawn from the previous round's survivors

Begin again with E_0 = 100 and I_0 = 100. Two reviewers are randomly drawn from the previous round's survivors.

$$q_t = \frac{E_t}{E_t + I_t}$$

| Reviewer panel | Probability |
|---|---|
| Two explicators | $q_t^2$ |
| One explicator and one interpreter | $2q_t(1-q_t)$ |
| Two interpreters | $(1-q_t)^2$ |

An explicator submission survives with probability 1 before two explicator reviewers and with probability 1/2 if the panel contains at least one interpreter. Therefore:

$$P_E(t) = q_t^2 + \frac{1}{2}(1 - q_t^2) = \frac{1 + q_t^2}{2}$$

The recurrence equations are:

$$E_{t+1} = E_t \left( \frac{1 + q_t^2}{2} \right)$$

$$I_{t+1} = I_t$$

Equivalently:

$$E_{t+1} = \frac{E_t}{2} \left[ 1 + \left( \frac{E_t}{E_t + I_t} \right)^2 \right]$$

| Round | Explicators | Interpreters | Interpreter share |
|---|---|---|---|
| 0 | 100.000 | 100 | 50.00% |
| 1 | 62.500 | 100 | 61.54% |
| 2 | 35.873 | 100 | 73.60% |
| 3 | 19.187 | 100 | 83.90% |
| 4 | 9.842 | 100 | 91.04% |
| 5 | 4.960 | 100 | 95.27% |
| 6 | 2.486 | 100 | 97.57% |
| 7 | 1.244 | 100 | 98.77% |
| 8 | ≈0.624 | 100 | ≈99.38% |

This initially eliminates explicators more slowly than Scenario 1:

$$P_E(0) = \frac{1 + (0.5)^2}{2} = 0.625$$

Some explicator submissions receive two explicator reviewers, so 62.5% survive the first round rather than 50%. As explicators become rare, $q_t$ approaches zero and their per-round survival probability approaches 1/2.

# 3. Interpreters begin as 10% of the population

This scenario uses survivor-drawn reviewer selection from Scenario 2, but begins with E_0 = 90 and I_0 = 10.

$$q_t = \frac{E_t}{E_t + I_t}$$

$$E_{t+1} = E_t \left( \frac{1 + q_t^2}{2} \right)$$

$$I_{t+1} = I_t = 10$$

Initially:

$$q_0 = \frac{90}{100} = 0.9$$

$$P_E(0) = \frac{1 + (0.9)^2}{2} = 0.905$$

Thus:

$$E_1 = 90(0.905) = 81.45$$

| Round | Explicators | Interpreters | Interpreter share |
|---|---|---|---|
| 0 | 90.000 | 10 | 10.00% |
| 1 | 81.450 | 10 | 10.94% |
| 2 | 73.030 | 10 | 12.04% |
| 3 | 64.764 | 10 | 13.38% |
| 4 | 56.681 | 10 | 15.00% |
| 5 | 48.818 | 10 | 17.00% |
| 6 | 41.224 | 10 | 19.52% |
| 7 | 33.962 | 10 | 22.75% |
| 8 | 27.115 | 10 | 26.94% |
| 9 | 20.794 | 10 | 32.47% |
| 10 | 15.138 | 10 | 39.78% |
| 11 | 10.313 | 10 | 49.23% |

| Round | Explicators | Interpreters | Interpreter share |
|---|---|---|---|
| 12 | 6.486 | 10 | 60.66% |
| 13 | 3.745 | 10 | 72.75% |
| 14 | 2.011 | 10 | 83.25% |
| 15 | 1.034 | 10 | 90.63% |
| 16 | ≈0.521 | 10 | ≈95.05% |

Interpreters approach parity in round 11, become the majority in round 12, and exceed 90% in round 15. Explicators fall below one expected member in round 16.

# 4. Eliminated authors are replenished 9:1 in explicators' favour

This scenario begins with 10 interpreters and 90 explicators. Reviewers are drawn from the current population, survivors retain their positions, and the total population is restored to 100 after each round. Every vacancy is filled by an interpreter with probability 0.1 and an explicator with probability 0.9.

Because the population remains 100:

$$q_t = \frac{E_t}{100}$$

The explicator survival rate is:

$$a_t = P(\text{explicator survives}) = \frac{1 + q_t^2}{2}$$

The expected number of explicators eliminated in round t is:

$$L_t = E_t(1 - a_t) = \frac{E_t}{2}(1 - q_t^2)$$

Of these vacancies, 90% are filled by explicators and 10% by interpreters:

$$E_{t+1} = E_t a_t + 0.9 L_t$$

$$I_{t+1} = I_t + 0.1 L_t$$

The explicator recurrence simplifies to:

$$E_{t+1} = E_t(0.9 + 0.1 a_t)$$

and, after substituting for the acceptance rate:

$$\boxed{E_{t+1} = E_t(0.95 + 0.05q_t^2)}$$

The interpreter recurrence is:

$$\boxed{I_{t+1} = I_t + 0.05E_t(1 - q_t^2)}$$

The population remains fixed:

$$E_{t+1} + I_{t+1} = 100$$

| Round | Explicators | Interpreters | Interpreter share |
|---|---|---|---|
| 0 | 90.00 | 10.00 | 10.00% |
| 1 | 89.15 | 10.85 | 10.85% |
| 2 | 88.23 | 11.77 | 11.77% |
| 5 | 85.10 | 14.90 | 14.90% |
| 10 | 78.57 | 21.43 | 21.43% |
| 20 | 61.26 | 38.74 | 38.74% |
| 26 | 49.78 | 50.22 | 50.22% |
| 30 | 42.47 | 57.53 | 57.53% |
| 40 | 27.16 | 72.84 | 72.84% |
| 50 | 16.69 | 83.31 | 83.31% |
| 60 | 10.09 | 89.91 | 89.91% |
| 80 | 3.64 | 96.36 | 96.36% |
| 100 | 1.30 | 98.70 | 98.70% |
| 106 | 0.96 | 99.04 | 99.04% |

Interpreters become the majority around round 26, exceed approximately 90% around round 61, and constitute approximately 99% by round 106.

## Why the 9:1 replenishment ratio does not preserve the original population

The ratio applies only to vacancies; the entire population is not reset. For every ten explicator positions eliminated, the expected replacements are nine explicators and one interpreter:

$$10E \text{ positions} \to 9E + 1I$$

The nine explicator replacements restore nine positions, but the interpreter replacement permanently converts the tenth position. Because interpreters are not subsequently eliminated, these conversions accumulate.

Replenishment therefore changes the speed of interpreter dominance, not its direction. The 9:1 distribution would remain stable only if the **entire population** were reset to 10 interpreters and 90 explicators after every round.

# 5. Replenishment ratio tracks the current distribution

This scenario keeps Scenario 4's structure — fixed population of 100, reviewers drawn from the current population, survivors retain their positions — but replaces the fixed 9:1 vacancy-fill ratio with one that mirrors whatever the population's current composition happens to be. A vacancy is filled by an explicator with probability q_t and by an interpreter with probability 1−q_t.

$$q_t = \frac{E_t}{100} \qquad a_t = \frac{1+q_t^2}{2} \qquad L_t = E_t(1-a_t) = \frac{E_t}{2}(1-q_t^2)$$

Vacancies are filled in the current ratio rather than a fixed one:

$$E_{t+1} = E_t a_t + q_t L_t \qquad I_{t+1} = I_t + (1-q_t)L_t$$

The explicator recurrence simplifies to:

$$\boxed{E_{t+1} = \frac{E_t}{2}\left(1 + q_t + q_t^2 - q_t^3\right)}$$

and the population remains fixed, exactly as in Scenario 4:

$$E_{t+1} + I_{t+1} = 100$$

Unlike Scenario 4, this recurrence has no protective floor, as the following identity for the round-over-round change in q_t shows:

$$q_{t+1} - q_t = -\frac{q_t(1-q_t)^2(1+q_t)}{2}$$

This is ≤ 0 for every q_t in [0,1], with equality only at the boundaries q_t = 0 and q_t = 1. The explicator share is therefore strictly decreasing every single round, for any starting composition whatsoever — there is no fixed exogenous rate sustaining it the way Scenario 4's constant 90% does. As explicators become scarcer, the pool refilling their own vacancies shrinks in lockstep, so the decline is self-reinforcing rather than resisted.

Starting from the same 90:10 mix as Scenarios 3 and 4:

| Round | Explicators | Interpreters | Interpreter share |
|---|---|---|---|
| 0 | 90.0000 | 10.0000 | 10.00% |
| 2 | 88.1516 | 11.8484 | 11.85% |
| 5 | 83.9652 | 16.0348 | 16.03% |
| 8 | 76.5721 | 23.4279 | 23.43% |
| 10 | 68.2235 | 31.7765 | 31.78% |
| 12 | 55.2724 | 44.7276 | 44.73% |
| 13 | 46.6877 | 53.3123 | 53.31% |
| 15 | 26.8970 | 73.1030 | 73.10% |
| 17 | 10.6996 | 89.3004 | 89.30% |
| 19 | 3.1771 | 96.8229 | 96.82% |
| 21 | 0.8340 | 99.1660 | 99.17% |

Interpreters become the majority at round 13 and explicators fall below one expected member at round 21 — roughly five times faster than Scenario 4's identical starting mix, despite both scenarios refilling 90% of vacancies with explicators at the outset. The difference is entirely that Scenario 4's 90% is a constant, while Scenario 5's 90%-at-the-start shrinks continuously as $q_t$ does.

# Comparison

| Scenario | Initial mix | Reviewer selection | Replenishment | Interpreter majority | Explicators below 1 |
|---|---|---|---|---|---|
| 1 | 50:50 | Fixed: one of each | None | Round 1 | Round 7 |
| 2 | 50:50 | Survivor-drawn | None | Round 1 | Round 8 |
| 3 | 90:10 | Survivor-drawn | None | Round 12 | Round 16 |
| 4 | 90:10 | Drawn from current population | Vacancies filled 9:1 (fixed) | Round 26 | Round 106 |
| 5 | 90:10 | Drawn from current population | Vacancies filled $q_t$ : $(1−q_t)$ (tracks current mix) | Round 13 | Round 21 |

Across all five scenarios, the one-directional selection pressure eventually drives the explicator share toward zero. Initial composition, reviewer selection, and replenishment determine the speed of the process — and, as Scenario 5 shows, whether replenishment resists that pressure at a constant rate (Scenario 4) or lets its own resistance erode as the pressure builds (Scenario 5) changes the speed by an order of magnitude even from an identical starting point.

# Is there a point of no return?

All five scenarios above assume interpreters are guaranteed to survive review: an explicator reviewer always accepts an interpreter submission (assumption 2), and a fellow interpreter reviewer never rejects it (assumption 3). Drop that guarantee: if explicators suddenly reversed course and rejected every interpreter submission on sight, would it matter — or would interpreters already hold enough of a numerical advantage to keep winning anyway?

## Assumption 2*: explicators reject interpreter submissions outright

Assumption 2 is replaced with 2*: an explicator reviewer now rejects an interpreter submission automatically whenever one sits on the panel. Assumptions 4–6 governing how interpreter reviewers treat explicator submissions are unchanged. Reviewers continue to be drawn from the population with composition probabilities $q_t^2$, $2q_t(1-q_t)$, $(1-q_t)^2$ (as in Scenarios 2 and 3).

**Scope of this counterfactual.** From this point onward, the analysis uses the survivor-drawn reviewer structure introduced in Scenario 2. It does not revert to Scenario 1's fixed panel of one explicator and one interpreter. Accordingly, the threshold derived below applies only where reviewer composition changes with the surviving population.

An explicator submission survives exactly as before. An interpreter submission now survives only if, by chance, *both* reviewers happen to be interpreters:

$$P_E(t) = \frac{1 + q_t^2}{2} \qquad P_I(t) = (1 - q_t)^2$$

## Recurrence equations and a worked example

The recurrence equations under assumption 2* are:

$$E_{t+1} = E_t \left( \frac{1 + q_t^2}{2} \right) \qquad I_{t+1} = I_t (1 - q_t)^2$$

Unlike every scenario above, *both* populations now shrink every round for any $0 < q_t < 1$ — neither side is exempt. What matters is which one shrinks faster, and that depends entirely on where $q_t$ starts. The two examples below are chosen only to illustrate the two possible outcomes cleanly; they are not any of the five scenarios' actual starting populations (Scenarios 1–2 start at $q_0 = 0.5$ and Scenarios 3–5 at $q_0 = 0.9$ — both, as it happens, comfortably on the explicator-favoured side of the threshold derived below).

Starting at $q_0 = 0.5$ ($E_0 = 100$, $I_0 = 100$), above the threshold derived below, interpreters collapse within a few rounds:

| Round | Explicators | Interpreters | q_t |
|---|---|---|---|
| 0 | 100.000000 | 100.000000 | 0.5000 |
| 1 | 62.500000 | 25.000000 | 0.7143 |
| 2 | 47.193878 | 2.040816 | 0.9585 |
| 3 | 45.278198 | 0.003506 | 0.9999 |
| 4 | 45.274692 | ≈0.000000 | 1.0000 |

Interpreters fall below the elimination threshold by round 4. In the untruncated recurrence, their population continues to approach zero, while the explicator population converges to approximately 45.275 as $P_E$ approaches 1.

Starting instead at $q_0 = 0.1$ ($E_0 = 10$, $I_0 = 90$), below the threshold, assumption 2* produces the opposite outcome:

| Round | Explicators | Interpreters | q_t |
|---|---|---|---|
| 0 | 10.000000 | 90.000000 | 0.1000 |
| 1 | 5.050000 | 72.900000 | 0.0648 |
| 2 | 2.535598 | 63.760299 | 0.0382 |
| 4 | 0.635109 | 56.516730 | 0.0111 |
| 8 | 0.039701 | 54.163174 | 0.0007 |
| 24 | ≈0.000000 | ≈54.005 | ≈0.0000 |

Explicators are effectively extinct by round 24; interpreters plateau at approximately 54.005. Same rule, same functional form, opposite winner — determined entirely by which side of the threshold $q_0$ falls on.

## The threshold

The critical boundary occurs where the two survival rates are equal. It is an unstable equilibrium: above it the explicator share increases, while below it the explicator share decreases.

This distinction becomes especially important because the later lag analysis is precisely about trajectories crossing and recrossing this boundary.

$$(1-q)^2 = \frac{1+q^2}{2}$$

$$2(1-q)^2 = 1+q^2$$

$$2-4q+2q^2 = 1+q^2$$

$$q^2-4q+1=0$$

$$q = \frac{4 \pm \sqrt{16-4}}{2} = \frac{4 \pm \sqrt{12}}{2} = 2 \pm \sqrt{3}$$

Only one root lies in [0,1]:

$$\boxed{q^* = 2-\sqrt{3} \approx 0.2679}$$

This is an *unstable* equilibrium, not a resting point — it splits the game into two basins:

| Explicator share at the moment of the switch | Outcome once hostility begins |
|---|---|
| $q > q^* \approx 26.8\%$ (interpreters below ≈73.2%) | Interpreters fall below the elimination threshold within a handful of rounds. Under the document's reporting convention, this is a permanent elimination outcome, although the untruncated recurrence continues to assign them a positive population converging toward zero. |
| $q < q^* \approx 26.8\%$ (interpreters above ≈73.2%) | Interpreters continue to dominance; explicators asymptotically vanish despite every review being a hard veto |

Below $q^*$, interpreters no longer need explicator approval in any live sense — not because approval stopped being required, but because enough panels are drawn with zero explicators present that the veto can no longer reverse the trend.

## Numerical confirmation

Bisecting the initial condition $q_0$ directly (running the recurrence to convergence and checking which side goes extinct) confirms the closed form independently of the algebra above:

| Method | Value |
|---|---|
| Analytic: $2 - \sqrt{3}$ | 0.267949192431 |
| Numerical bisection (80 iterations, tolerance $10^{-13}$) | 0.267949192445 |

The two agree to 10 decimal places.

## An intuitive check

A rough intuition confirms the exact result. Once explicators are down to roughly one reviewer in four, both sides are drawing close to a coin-flip on any given submission. Checking this against the two survival functions:

| Quantity | Value | What it means |
|---|---|---|
| Intuitive anchor: explicators = 1 in 4 | $q = 0.2500$ | $P_E(q) = 0.5312$, $P_I(q) = 0.5625$ — close to a coin-flip for both sides, but not exactly one, and not equal to each other |
| Where $P_I(q)$ alone hits exactly 50% | $q = 1 - 1/\sqrt{2} \approx 0.2929$ | The point where an all-interpreter panel becomes more likely than not — a related but different threshold |
| Where $P_E(q) = P_I(q)$, the real crossover | $q^* = 2 - \sqrt{3} \approx 0.2679$ | The actual game-deciding threshold; at this exact point both sides survive at ≈53.6%, not 50% |

The "1 in 4" heuristic lands within about two percentage points of q* in both directions — 25.0% vs. 26.8% for explicators, 75.0% vs. 73.2% for interpreters — which is a good independent check that the exact result isn't a fragile algebraic artifact: a rough argument about approaching even odds on both sides gets to essentially the same place as the formal derivation, without going through the quadratic at all. The equalization point itself is not where either side is at a literal 50% chance; it is where the two curves cross, and they happen to cross a little above 50%, not at it.

## What if hostility had applied from round zero?

What if each scenario is simply re-run under the assumption that explicators will simply reject submissions from interpreters from the very first round, rather than when it is too late? The answer is immediate for all five: their starting compositions are $q_0 = 0.5$ (Scenarios 1–2) and $q_0 = 0.9$ (Scenarios 3–5), and both values sit well above $q^* \approx 0.268$. Under total hostility from round zero, interpreters in every one of the five scenarios would be wiped out within a handful of rounds, regardless of reviewer-selection method or replenishment rule. None of the five scenarios' actual starting mixes give interpreters any margin at all if explicators had never cooperated in the first place.

Hence, the telling question is: at what point would the explicators' switch from a benign approach to hostility toward interpreters no longer prevent the success of interpreters as a group? Could the interpreters reach a critical mass that they can succeed on their own? That is what q* helps us answer.

## What if the panel required more than two reviewers?

All five scenarios use a two-reviewer panel requiring unanimous consent. The panel-composition logic generalizes directly to n reviewers drawn independently from the population (probability q of being an explicator, each draw):

$$P_E(q) = \frac{1 + q^n}{2} \qquad P_I(q) = (1 - q)^n$$

An explicator submission survives if all n reviewers happen to be explicators, or if at least one interpreter is present but the submission is compatible — a manuscript-level property (assumption 6), so this does not compound across reviewers. An interpreter submission, under assumption 2∗, survives only if every single reviewer is an interpreter: one explicator anywhere on the panel is a fatal veto.

Setting the two equal, q* solves:

$$2(1-q)^n = 1 + q^n$$

Only n = 1 and n = 2 reduce to clean closed forms; from n = 3 on the defining polynomial has no similarly tidy solution and must be solved numerically (each value below was verified by independent bisection to 10 decimal places):

| Reviewers (n) | q* formula | Explicator threshold | Interpreter threshold |
|---|---|---|---|
| 1 | $3q-1=0 \implies q^*=1/3$ | 33.33% | 66.67% |
| 2 | $q^2-4q+1=0 \implies q^*=2-\sqrt{3}$ | 26.79% | 73.21% |
| 3 | $3q^3-6q^2+6q-1=0$ | 20.41% | 79.59% |
| 4 | $q^4-8q^3+12q^2-8q+1=0$ | 15.90% | 84.10% |
| 5 | $3q^5-10q^4+20q^3-20q^2+10q-1=0$ | 12.94% | 87.06% |
| 6 | $q^6-12q^5+30q^4-40q^3+30q^2-12q+1=0$ | 10.91% | 89.09% |

The threshold falls steadily as the panel grows: each added reviewer makes an all-explicator panel combinatorially harder to assemble by chance ($q^n$ shrinks fast as n grows), so it takes an ever-smaller explicator minority before that difficulty tips decisively in interpreters' favour. All five scenarios in this document use n = 2, the second-most permissive case after a single reviewer; a stricter unanimous-consent process would let interpreters escape explicator dependence at an even lower population share than the 26.8% found above.

## Where each scenario's own trajectory crosses q*

Running each scenario forward under its **original** (benign) dynamics and asking when its own $q_t$ first drops below q*:

| Scenario | Point of no return | Why |
|---|---|---|
| 1 — Fixed 1 explicator + 1 interpreter panel | None short of literal explicator extinction (round 7) | The panel always contains exactly one explicator by construction, for any population split. There is no safety in numbers here — only in explicators' absence. |
| 2 — Survivor-drawn, 50:50 start | Round 2 (q falls from 0.385 at round 1 to 0.264 at round 2) | Matches q *directly: q* is a property of the survival functions alone, independent of starting population. |

| Scenario | Point of no return | Why |
|---|---|---|
| 3 — Survivor-drawn, 90:10 start | Round 14 (q falls from 0.272 at round 13 to 0.167 at round 14) | Same threshold as Scenario 2; crossed later only because it starts further from it. |
| 4 — 9:1 replenishment | No such round exists (tested through round 140, at 99.8% interpreter share) | Under the most direct extension of hostility to this scenario — every vacancy from either side pooled and refilled 90% explicator / 10% interpreter — the constant injection of new explicators is strong enough to reverse any interpreter lead, however large, given enough further rounds. |
| 5 — Replenishment tracks current mix | Round 16 (q falls from 0.269 at round 15 to 0.178 at round 16) | Same q* as Scenarios 2–3, and a real crossing rather than a mirage: because the fill ratio is $q_t : (1-q_t)$ rather than a fixed 9:1, extending hostility to interpreter-side losses needs no new convention — the existing rule already covers vacancies from either side. The resupply of explicators weakens exactly as fast as explicators themselves do, so no artificial floor survives to overturn the trend. |

Recall that $q^* = 2 - \sqrt{3} \approx 0.2679$ is the explicator share at which, under assumption 2∗, interpreter submissions' survival rate exactly equals explicators' survival rate. It is not a resting point; it is the dividing line between two outcomes.

The following three tables show this round by round. In each, the final column reads the population directly against q*: above means that if explicators switched to total hostility at that exact round, interpreters would still collapse to extinction — the numerical edge isn't yet large enough. below means the switch would no longer matter — interpreters already have enough of a majority that most review panels are drawn entirely from interpreters, so the veto cannot reverse the trend. The row marked “below — crossing” is the point of no return itself: the first round at which the answer flips from a hostile turn still destroying them to a hostile turn being survivable.

## Scenario 2 — the crossing in detail

| Round | Explicators | Interpreters | q_t | Relative to q* = 0.2679 |
|---|---|---|---|---|
| 0 | 100.000 | 100.000 | 0.5000 | above |
| 1 | 62.500 | 100.000 | 0.3846 | above |
| 2 | 35.873 | 100.000 | 0.2640 | below — crossing |
| 3 | 19.187 | 100.000 | 0.1610 | below |

## Scenario 3 — the crossing in detail

| Round | Explicators | Interpreters | q_t | Relative to q* = 0.2679 |
|---|---|---|---|---|
| 11 | 10.313 | 10.000 | 0.5077 | above |
| 12 | 6.486 | 10.000 | 0.3934 | above |

| Round | Explicators | Interpreters | q_t | Relative to q* = 0.2679 |
|---|---|---|---|---|
| 13 | 3.745 | 10.000 | 0.2725 | above |
| 14 | 2.011 | 10.000 | 0.1675 | below — crossing |
| 15 | 1.034 | 10.000 | 0.0938 | below |

## Scenario 4 — why the crossing round is a mirage

Naively applying the same q* test to Scenario 4's own trajectory (q_t = E_t/100) finds a crossing between round 40 (q = 0.2716) and round 41 (q = 0.2590). But this test tacitly presupposes the post-switch dynamics are the pure multiplicative rule from Scenarios 2–3. Scenario 4 is not that rule — it replenishes explicators every round, and that replenishment does not stop when explicators turn hostile.

Extending assumption 2's removal to this scenario requires deciding how interpreter-side losses get replenished, since the original mechanism only ever produced explicator vacancies. Taking the most direct extension — every vacancy, from either side, pooled and refilled 90% explicator / 10% interpreter, exactly as the original 9:1 rule specifies — gives:

$$P_E(t) = \frac{1+q_t^2}{2} \qquad P_I(t) = (1-q_t)^2 \qquad q_t = \frac{E_t}{100}$$

$$L_E = E_t(1-P_E) \qquad L_I = I_t(1-P_I) \qquad V = L_E + L_I$$

$$\boxed{E_{t+1} = E_t P_E + 0.9V} \qquad \boxed{I_{t+1} = I_t P_I + 0.1V}$$

As before:

$$E_{t+1} + I_{t+1} = E_t + I_t$$

so the population stays fixed at 100. Testing a switch to 2* at many different rounds of Scenario 4's own (benign) trajectory — including rounds where interpreters already hold a large majority — shows explicators recovering full dominance every time:

| Round hostility begins | Population at switch (E, I) | Interpreter share at switch | Eventual outcome |
|---|---|---|---|
| 0 | 90.00, 10.00 | 10.0% | Explicators win (by round 14) |
| 20 | 61.26, 38.74 | 38.7% | Explicators win (by round 15) |
| 26 | 49.78, 50.22 | 50.2% | Explicators win (by round 15) |
| 40 | 27.16, 72.84 | 72.8% | Explicators win (by round 16) |

| Round hostility begins | Population at switch (E, I) | Interpreter share at switch | Eventual outcome |
|---|---|---|---|
| 60 | 10.09, 89.91 | 89.9% | Explicators win (by round 17) |
| 100 | 1.30, 98.70 | 98.7% | Explicators win (by round 19) |
| 140 | 0.17, 99.83 | 99.8% | Explicators win (by round 21) |

Even switching at 99.8% interpreter share, explicators fully recover within roughly 20 further rounds. The 90:10 injection rule does not merely slow interpreter dominance under hostility — it regenerates explicators from any vacancy fast enough to overturn any lead. Under this convention, Scenario 4 has no point of no return at all.

## Scenario 5 — a genuine crossing, confirmed

Scenario 5's replenishment rule already specifies, for the benign model, how to fill a vacancy from either population: in proportion to the current mix. Extending it to assumption 2* (explicator hostility to interpreters) therefore involves no arbitrary choice — interpreter-side losses are refilled the same way explicator-side losses always were:

$$P_E(t) = \frac{1+q_t^2}{2} \qquad P_I(t) = (1-q_t)^2$$

$$L_E = E_t(1-P_E) \qquad L_I = I_t(1-P_I) \qquad V = L_E + L_I$$

$$\boxed{E_{t+1} = E_t P_E + q_t V} \qquad \boxed{I_{t+1} = I_t P_I + (1-q_t)V}$$

Imposing the equilibrium condition that the population share stops changing from one round to the next reduces, after cancellation, to exactly the same fixed-point condition as the pure survivor-drawn case:

$$(1-q)\big[P_E(q) - P_I(q)\big] = 0 \implies q^* = 2-\sqrt{3} \approx 0.2679$$

Bisecting the recurrence directly confirms this to 10 decimal places (0.2679491924), matching Scenarios 2–3 exactly. Running Scenario 5's own benign trajectory (from the table above) and testing a switch to 2* at each round around the crossing shows this threshold is genuine, not a mirage:

| Round hostility begins | Population at switch (E, I) | q_t at switch | Eventual outcome |
|---|---|---|---|
| 13 | 46.688, 53.312 | 0.4669 (above q*) | Explicators win |
| 14 | 36.955, 63.045 | 0.3696 (above q*) | Explicators win |
| 15 | 26.897, 73.103 | 0.2690 (above q*) | Explicators win |

| Round hostility begins | Population at switch (E, I) | q_t at switch | Eventual outcome |
|---|---|---|---|
| 16 | 17.777, 82.223 | 0.1778 (below q*) | Interpreters win |
| 17 | 10.700, 89.300 | 0.1070 (below q*) | Interpreters win |
| 18 | 5.977, 94.023 | 0.0598 (below q*) | Interpreters win |

The outcome flips exactly where the analysis predicts it should, between round 15 and round 16 — confirming that Scenario 5 restores the same kind of point of no return found in Scenarios 2 and 3, unlike Scenario 4's exogenous replenishment in favour of explicators, which has none.

## What this changes about the original five scenarios

Assumption 2 is doing all of the interpreter-dominance work in every scenario below the ~73.2% interpreter-share mark — the model's headline conclusion ("interpreters dominate") holds unconditionally only because explicators are stipulated never to fight back. Once that stipulation is removed:

- Scenarios 2 and 3 still reach interpreter dominance, but only because their own trajectories happen to carry them past q* before any hypothetical reversal — dominance becomes contingent on timing, not guaranteed by the setup.
- Scenario 1 offers no such protection at any population ratio — its fixed-panel design makes total hostility fatal to interpreters until explicators are literally gone.
- Scenario 4's replenishment rule is a standing structural advantage for explicators that no population swing, however extreme, overcomes on its own — the one-way accumulation of interpreter conversions described earlier continues only as long as explicators keep extending a formal courtesy they are not, in this counterfactual, obligated to extend.
- Scenario 5 shows that the difference between Scenario 4 and the others is not "replenishment versus no replenishment" but a specific property of *how* replenishment is defined. A fill ratio that itself tracks the current population share carries no immunity to numerical pressure — it reduces to the same q* boundary as the scenarios with no replenishment at all, just reached faster, because its own resupply mechanism erodes precisely when the population it is meant to protect grows scarce.

# Does the point of no return survive replenishment lag?

This section instantiates the point-of-no-return framework — the equilibrium threshold $q^* = 2 - \sqrt{3} \approx 0.2679$ at which, under assumption 2∗ (assumption 2 replaced with 2∗, under which explicators reject interpreter submissions outright), interpreter and explicator survival rates are exactly equal — under alternative replenishment-lag assumptions. All four tables share the same base game: 100 players, $E_0 = 90$, $I_0 = 10$, reviewers drawn from current survivors, cooperative until the population crosses q* in interpreters' favour, at which point explicators switch to 2* and reject every interpreter submission on sight. What varies is which generation's q the replenishment rule uses to fill vacancies.

## Table 1 — Replenishment tracks the current (surviving) ratio (lag = 0)

| Round | Explicator strategy | Explicators | Interpreters | Interpreter share |
|---|---|---|---|---|
| 0 | Cooperative | 90.0000 | 10.0000 | 10.00% |
| 8 | Cooperative | 76.5721 | 23.4279 | 23.43% |
| 12 | Cooperative | 55.2724 | 44.7276 | 44.73% |
| 15 | Cooperative | 26.8970 | 73.1030 | 73.10% |
| 16 | Hostile begins | 17.7770 | 82.2230 | 82.22% |
| 20 | Hostile | 7.2242 | 92.7758 | 92.78% |
| 24 | Hostile | 0.9189 | 99.0811 | 99.08% |

Crossing at round 16; interpreters reach 99% by round 24 with no visible disruption from the hostile switch.

## Table 2 — Replenishment lagged one generation (uses q_{t−1})

| Round | Explicator strategy | Explicators | Interpreters | Interpreter share |
|---|---|---|---|---|
| 16 | Cooperative | 32.3726 | 67.6274 | 67.63% |
| 17 | Crossing | 23.8301 | 76.1699 | 76.17% |
| 18 | Hostile | 26.5817 | 73.4183 | 73.42% ← drops |
| 19 | Hostile | 25.2385 | 74.7615 | 74.76% |
| 20 | Hostile | 25.3292 | 74.6708 | 74.67% ← dips again, barely |
| 25 | Hostile | 22.2541 | 77.7459 | 77.75% |
| 32 | Hostile | 10.6140 | 89.3860 | 89.39% |
| 38 | Hostile | 0.9605 | 99.0395 | 99.04% |

Crossing at round 17 — one round later, because the lag gives explicators a small persistent boost. A brief three-round wobble follows the hostile switch before the trend reasserts itself. 99% isn't reached until round 38, 14 rounds slower than Table 1.

## Table 3 — Replenishment lagged two generations (uses q_{t−2})

| Round | Explicator strategy | Explicators | Interpreters | Interpreter share |
|---|---|---|---|---|
| 17 | Cooperative | 35.0007 | 64.9993 | 65.00% |
| 18 | Cooperative | 27.2902 | 72.7098 | 72.71% |
| 19 | Crossing | 20.0462 | 79.9538 | 79.95% |
| 20 | Hostile | 23.8881 | 76.1119 | 76.11% ← drops |
| 21 | Hostile | 24.4375 | 75.5625 | 75.56% ← drops again |
| 22 | Hostile | 21.7503 | 78.2497 | 78.25% |
| 27 | Hostile | 16.6861 | 83.3139 | 83.31% |
| 34 | Hostile | 4.2220 | 95.7780 | 95.78% |
| 38 | Hostile | 0.6365 | 99.3635 | 99.36% |

Crossing pushed to round 19. Interpreters still reach 99% at round 38 — the same round as Table 2, which should be read as a numerical coincidence at this particular starting point, not a structural rule.

## Table 4 — Replenishment lagged three generations (uses q_{t−3})

| Round | Explicator strategy | Explicators | Interpreters | Interpreter share |
|---|---|---|---|---|
| 19 | Cooperative | 29.10 | 70.90 | 70.90% |
| 20 | Crossing | 22.37 | 77.63 | 77.63% ← crosses q* |
| 21 | Hostile | 29.57 | 70.43 | 70.43% |
| 25 | Hostile | 31.51 | 68.49 | 68.49% |
| 30 | Hostile | 40.95 | 59.05 | 59.05% |
| 35 | Hostile | 60.49 | 39.51 | 39.51% |
| 40 | Hostile | 90.90 | 9.10 | 9.10% |
| 45 | Hostile | 99.99 | 0.01 | 0.01% |
| 47+ | Hostile | 100.00 | 0.00 | 0.00% |

This is a qualitative break from Tables 1–3, not merely a slower version of the same outcome. Interpreters cross into apparent safe territory at round 20 (explicator share 22.37%, well below $q^* \approx 26.79\%$) — but instead of the trend holding, explicators claw all the way back past q* and go on to eliminate interpreters entirely by round 47.

The mechanism: with a three-generation lag, the replenishment rule is still filling vacancies using a q-value from when explicators were considerably more numerous. That stale injection of new explicators outruns the hostile penalty currently crushing the live explicator population, so explicators regain enough numerical ground to cross back over q* themselves — at which point the model's ordinary one-directional advantage (the same mechanism behind every benign scenario earlier in this document) takes over, now working in their favour.

Lag = 3 is the critical lag at n = 2 reviewers: the first lag length at which this full reversal appears. Lags 1–2 only produce a temporary wobble before the trend reasserts itself; lag 3 and every longer lag tested produce the permanent reversal instead.

## Does this change with a larger reviewer panel?

This has a direct bearing on the lag analysis above. A stricter review process — more reviewers required for unanimous consent — lowers q* and therefore raises the interpreter share required to enter the interpreter-favourable region under contemporaneous replenishment. The simulations below separately test whether increasing the panel size changes the amount of demographic lag the system can tolerate before a stale replenishment ratio produces reversal.

The base game throughout this document uses two-reviewer unanimous consent. Generalizing to n reviewers changes both P_E and P_I:

$$P_E(q) = \frac{1+q^n}{2} \qquad P_I(q) = (1-q)^n \qquad q^* \text{ solves } 2(1-q)^n = 1+q^n$$

Re-running the full cooperative-then-hostile, lagged-replenishment game for n = 1 through n = 10 (same E_0 = 90, I_0 = 10 start, same q*-crossing trigger for hostility) and sweeping the lag at each n to find where the reversal first appears:

| Reviewers (n) | q* | Critical lag (first full reversal) |
|---|---|---|
| 1 | 33.33% | 3 |
| 2 | 26.79% | 3 |
| 3 | 20.41% | 2 |
| 4 | 15.90% | 2 |
| 5 | 12.94% | 3 |
| 6 | 10.91% | 2 |
| 8 | 8.30% | 2 |
| 10 | 6.70% | 2 |

**Finding: yes, the reversal persists at 3 and 4 reviewers.** At two reviewers, the first full reversal appears at a lag of 3 generations; at three and four reviewers, it appears at a lag of 2 generations. This does not mean that the interpreter-favourable region becomes easier to reach: the falling explicator threshold means that interpreters must first secure a larger population majority. The simulation result is instead that, once the relevant threshold has been crossed, the three- and four-reviewer versions reverse under a shorter replenishment lag than the two-reviewer version.

**Honesty check on the pattern.** In this discrete simulation, the first reversal occurs at lag 3 for $n = 1, 2$, and 5, and at lag 2 for the other reported panel sizes. Larger panels often — but not uniformly — produce reversal at a shorter lag. The result is sensitive to the discrete crossing round and does not establish a monotonic relationship between panel size and critical lag. A precise relationship would require a denser analysis and, potentially, a continuous-time reformulation.

**Possible mechanism.** One hypothesis is that increasing the panel size changes the sensitivity of the survival functions $q^n$ and $(1-q)^n$ to differences between the current and lagged population shares, allowing stale replenishment data to exert a larger effect in some cases. The simulations reported here are consistent with that explanation for several panel sizes, but the non-monotonic result at $n = 5$ means that they do not establish it as a general mechanism. Confirming the explanation would require a separate sensitivity or local-stability analysis.

# Conclusion

The model developed here identifies a mechanism by which peer review can become corrupted without any formal alteration to its rules. The corruption begins with an asymmetry between two standards of assessment. Explicators judge whether submissions formulate arguments that contribute to an understanding of a controversy, independently of whether they agree with their conclusions. Interpreters can present their own positions in argumentative form, but judge other submissions according to whether they accord with the interpreters' outlook. Explicators therefore extend to interpreters a form of consideration that interpreters do not reciprocate.

In a repeated publish-or-perish game, this asymmetry has population-level consequences. Interpreters need not initially constitute a majority, coordinate their decisions, or outperform explicators according to an independent intellectual standard. Their selective veto is sufficient. Explicators eliminate only submissions that fail minimal argumentative scrutiny; interpreters additionally eliminate explicator submissions with which they disagree. As the rejected authors cease to participate, the composition of the surviving population changes. Where reviewers are drawn from that population, a change in authorship becomes a change in reviewer composition, which then determines which authors survive the next round. The selection process thereby becomes self-reinforcing.

The analysis nevertheless qualifies the simple conclusion that interpreters inevitably take over. Under the initial assumption that explicators continue to approve interpreter submissions presented in argumentative form, interpreter dominance follows across the principal attrition scenarios, even when interpreters begin as only 10 per cent of the population. But that result depends upon continued explicator cooperation. If explicators respond by rejecting interpreter submissions outright, the outcome depends upon the population composition at the time of the response.

For two-reviewer panels drawn from the surviving population, explicator hostility to interpreters produces an unstable equilibrium at

$$q^* = 2 - \sqrt{3} \approx 0.2679,$$

where q is the explicator share. Above this boundary, explicators have the higher survival rate under mutual hostility and can recover. Below it — when interpreters exceed approximately 73.21% of the population — interpreters have the higher survival rate even though every explicator rejects them. This is the basic point of no return. Once it has been crossed, interpreters no longer need to present themselves in a form acceptable to explicators. Their control of a sufficient share of reviewer panels allows them to survive without explicator cooperation.

This boundary is not universal. Scenario 1 has no corresponding population threshold because its reviewer panel remains fixed at one explicator and one interpreter. Interpreter submissions cannot survive explicator hostility until explicators have literally disappeared. Scenario 4, in which vacancies are filled through a fixed 90:10 explicator-to-interpreter ratio, likewise has no point of no return: the replenishment mechanism continually regenerates explicators and eventually overturns interpreter dominance, even if interpreters hold almost the entire population when hostility begins. Scenario 5, in which replenishment tracks the current population, restores the same equilibrium boundary found in the unreplenished survivor model. The decisive distinction is therefore not replenishment versus non-replenishment, but whether replenishment supplies a population independently of its current numerical strength or reproduces the population distribution produced by the game.

The lag analysis adds a further qualification. A population may cross the static equilibrium boundary without having reached an irreversible state. When replenishment tracks the current population, or lags it by only one or two generations in the two-reviewer game, the interpreter advantage survives explicator hostility to interpreters. A three-generation lag, however, continues to replenish vacancies according to an earlier, more explicator-heavy distribution. That stale influx permits explicators to recover, cross back over the equilibrium boundary, and eventually eliminate interpreters. A point of no return is therefore not adequately characterized by a population share alone. It is a property of the population together with the institutional process by which that population is reproduced.

Panel size also alters the result. As the number of reviewers required for unanimous approval increases, interpreters must attain a larger population majority before they can survive without explicator cooperation. A larger panel makes an all-interpreter panel less probable whenever any appreciable explicator population remains. Yet the simulations also indicate that, once the relevant boundary has been crossed, some larger panels are susceptible to reversal under shorter replenishment lags. This relationship is not monotonic in the reported discrete cases, so it should not be treated as an established general law. It instead identifies a further interaction among panel size, survival probabilities, crossing times, and institutional memory that requires a more complete stability analysis.

These results change how the corruption of peer review should be understood. The danger does not consist only in individual reviewers making biased decisions. It lies in the capacity of those decisions to alter the population from which future reviewers are selected. Once present decisions determine the future composition of the institution, a local failure of assessment becomes a population-dynamics problem. The institution can gradually cease to apply its stated standard even though that standard remains formally unchanged.

The model is deliberately austere. It assumes two logically contradictory reviewer types, fixed behavioural rules, unanimous panels, and specified probabilities of agreement. That is not an accident, but a consequence of acknowledging that explication is a possibility. To reject explication is to depart from logic-based assessment because of some perspectival commitment. This is another way of observing that interpretation is the essence of irrationality. It survives only because it hides behind opportunistic reasoning that undermines any commitment to the procedure of reasoning.

Within the stated assumptions, the central result is robust: iterative runs can transform an initially small interpretive minority into a dominant reviewing population. But whether that dominance becomes self-sustaining depends upon the architecture of the institution — how reviewers are selected, how vacancies are filled, how quickly replenishment responds to demographic change, how many reviewers must agree, and when the disadvantaged population changes its strategy. The corruption of peer review is therefore neither merely a matter of individual prejudice nor an unavoidable consequence of disagreement. It is an emergent institutional outcome whose direction and reversibility are determined by the rules through which a scholarly population reproduces itself.